\documentclass[twocolumn,showpacs,letterpaper,superscriptaddress]{revtex4}

\usepackage{graphicx,amsmath,amssymb}

\begin{document}

\newcommand{\ket}[1]{\left| #1 \right\rangle}
\newcommand{\bra}[1]{\left\langle #1\right |}
\newcommand{\indfirst}[2]{{\vphantom{#2}}_{#1}{#2}}
\def\mathbi#1{\textbf{\em #1}}

\title{Quantum phase gate and controlled entanglement with polar molecules}

\author{Eric Charron}
\affiliation{Laboratoire de Photophysique Mol\'eculaire du CNRS,\\
Universit\'e Paris-Sud, B\^atiment 210, 91405 Orsay Cedex, France}

\author{P\'erola Milman}
\affiliation{Laboratoire de Photophysique Mol\'eculaire du CNRS,\\
Universit\'e Paris-Sud, B\^atiment 210, 91405 Orsay Cedex, France}
\affiliation{CERMICS, Ecole Nationale des Ponts et Chauss\'ees,\\
6 et 8 av. Blaise Pascal, Cit\'e Descartes,\\
Champs-sur-Marne, 77455 Marne-la-Vall\'ee, France }

\author{Arne Keller}
\affiliation{Laboratoire de Photophysique Mol\'eculaire du CNRS,\\
Universit\'e Paris-Sud, B\^atiment 210, 91405 Orsay Cedex, France}

\author{Osman Atabek}
\affiliation{Laboratoire de Photophysique Mol\'eculaire du CNRS,\\
Universit\'e Paris-Sud, B\^atiment 210, 91405 Orsay Cedex, France}

\date{\today}

\begin{abstract}
We propose an alternative scenario for the generation of entanglement between rotational quantum states of two polar molecules. This entanglement arises from dipole-dipole interaction, and is controlled by a sequence of laser pulses simultaneously exciting both molecules. We study the efficiency of the process, and discuss possible experimental implementations with cold molecules trapped in optical lattices or in solid matrices. Finally, various entanglement detection procedures are presented, and their suitability for these two physical situations is analyzed.
\end{abstract}


\maketitle

\section{Introduction}

In the last years, the development of quantum information opened new perspectives for several physical systems displaying controllable quantum properties. While some low dimensional basic quantum information tools have been experimentally realized with cavity quantum electrodynamics\,\cite{HAROCHE}, trapped ions\,\cite{BLATT_WINE}, NMR\,\cite{NMR} and cold atoms\,\cite{NA}, for example, the exploitation of other systems presenting potential advantages remains of great importance. In this paper, we focus our attention to polar diatomic molecules.

There is a growing recent interest in exploiting molecules for quantum information purposes, both from the theoretical and experimental points of view. This interest comes partly from the development of new methods for the generation of ultracold molecular gases\,\cite{Revue_Masnou}. Two techniques are now widely used for this purpose: photoassociation\,\cite{PA} and magnetic Feshbach resonances\,\cite{FR}. These methods, first applied to the formation of homonuclear molecules, were then used for the creation of ultracold polar heteronuclear molecules such as RbCs\,\cite{RbCs}, LiNa\,\cite{LiNa}, KRb\,\cite{KRb} or NaCs\,\cite{NaCs} in various trapping situations. Very high formation efficiencies were also obtained recently for homonuclear molecules in optical lattices, when two atoms are located in each lattice site\,\cite{Mol_OL}. This kind of trap presents several advantages since it allows for the control of both the internal rovibronic and external center of mass quantum states of the molecules, which can additionally be isolated from each other due to the tight confinement at the lattice sites. Finally, the controlled creation of cold heteronuclear molecules in optical lattices\,\cite{H_Mol_OL} would allow for the production of strong inter-molecular interactions which could be exploited for quantum computation\,\cite{Revue_Masnou}.

Indeed, several proposals have been presented recently to benefit from the specificities of molecules for quantum information\,\cite{ERIC,MOLQI,MOLQI2,MOLQI3,ZOLLER}.
Theoretically, it was shown that molecules can be used to store binary information in the phases of rotational wavepackets\,\cite{ERIC}. The implementation of simple quantum algorithms has also been proposed using femtosecond pulses acting on diatomic molecules\,\cite{MOLQI,MOLQI2,MOLQI3}. Additionally, it was shown that molecules in optical lattices can simulate topological order, generating topologically protected subspaces where a quantum bit (qubit) can be encoded\,\cite{ZOLLER}.

The use of polar diatomic molecules in various kinds of traps was also proposed recently by different groups for quantum computation\,\cite{DEMILLE,LEE,KOTO,Yelin}. DeMille\,\cite{DEMILLE} first proposed to use molecules oriented along an external electric field in a 1D trap array for the implementation C-NOT gates with a very large number of qubits. Lee and Ostrovskaya\,\cite{LEE} then proposed to use coherent Raman transitions between scattering and bound states of heteronuclear molecules trapped in optical lattices for the implementation of conditional dipole-dipole interactions. It was shown by Kotochigova and Tiesinga\,\cite{KOTO} that microwave fields can be used to induce a tunable dipole-dipole interaction between ground state rotationally symmetric molecules. Finally, Yelin {\em et al\/}\,\cite{Yelin} also proposed very recently different schemes for the implementation of molecular quantum gates. From the experimental point of view and in another context, evidences of quantum correlations caused by dipole interaction between two molecules separated by tens of nanometers in an organic crystal were observed by detection of photon bunching\,\cite{SCIENCE}.

In the present work we address the question of controlled entanglement creation in cold polar diatomic molecular systems. In our proposal, pure states with any degree of entanglement can be created by laser assisted conditional dipole-dipole interaction. Compared to previous proposals, this process only involves two vibrational states in their three lowest rotational levels. A Raman transition is used to transfer the qubit state from a set of uninteracting levels used for the storage of information to a set of interacting levels used for the creation of entanglement. This ability to switch on and off the dipole interaction with simple optical pulses of relatively short duration allows for the implementation of conditional quantum logic. In this approach, the dipole interaction is sufficiently weak to be treated as a perturbation, but sufficiently strong to generate maximally entangled states in a relatively short duration \mbox{($\simeq 10\,\mu$s --} \mbox{1\,ms)} as compared to the expected coherence time \mbox{($\gg 1\,$s)}\,\cite{KOTO,KOTO2}. One could therefore perform about $~10^4$ logical gates within this anticipated coherence time. In addition to this possible implementation with cold molecules trapped in optical lattices, other physical systems, such as molecules trapped in solid matrices, are also considered in the present work. In this case, entanglement is created in an uncontrolled way in a collection of $N$ molecules, two by two. Finally, we investigate some ways of detecting entanglement given the possibilities of each system, and describe how to perform non-locality tests.

In Section\,\ref{sec:creation}, we describe the basic principles for entanglement creation between the rotational levels of two polar molecules. The effects of dissipation are discussed in Section\,\ref{sec:exp}, together with possible experimental realizations. Finally, a direct detection test of this entanglement is described in Section\,\ref{sec:detect}.

\section{Laser assisted creation of rotational entanglement}
\label{sec:creation}

\subsection{General frame}
\label{sec:GI}

We consider here two identical diatomic polar molecules, initially prepared in their ground electronic and rotational levels. For simplicity, and in order to describe the physical process on which relies this entanglement creation procedure, we ignore in this section the vibrational degree of freedom. The additional complexity introduced by the vibrational motion will be dealt with in Section\,\ref{sec:QPG}.

Since the mechanism proposed in the next Section for the implementation of a quantum phase gate is based on unitary transformations conserving the projection of the rotational quantum number of both molecules on the inter-molecular axis, this projection is fixed at zero in the following (see the justification given at the end of Sec.\,\ref{sec:QPG} for details).

\begin{figure}[!t]
\includegraphics[width=8.6cm,clip]{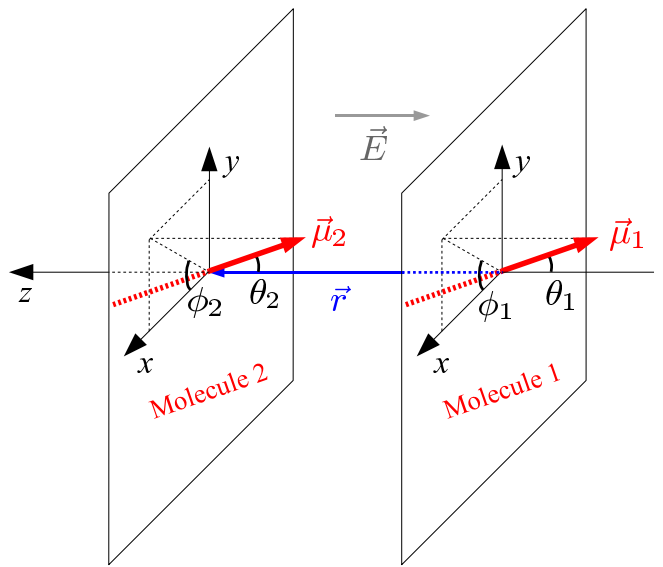}
\caption{(Color online) Schematic view of the molecular configuration. The quantization axis is chosen as the inter-molecular axis. The electric fields \mbox{$\vec{E}(t)$} associated with the laser pulses are assumed to be linearly polarized along this same direction. The orientations of the permanent dipoles $\vec{\boldsymbol{\mu}}_1$ and $\vec{\boldsymbol{\mu}}_2$ of the two molecules are characterized by the angles \mbox{$(\theta_1,\phi_1)$} and \mbox{$(\theta_2,\phi_2)$}.}
\label{fig1}
\end{figure}

The rotational stationary states of each isolated molecule, with energies
\begin{equation}
\label{En}
\varepsilon_N = B_{\mathrm{rot}}\, N(N+1)\,,
\end{equation}
are denoted by $\ket{N}_{i}$, with
\begin{equation}
\label{N}
\langle\theta_i,\phi_i|N\rangle_{i} = Y_{N,0}(\theta_i,\phi_i)\,,
\end{equation}
where $Y_{N,0}(\theta_i,\phi_i)$ represents the spherical harmonic associated with the molecular rotational quantum number $N$ of projection zero on the inter-molecular axis. The index identifying each molecule is \mbox{$i=1,2$}. The angular coordinates of the two molecules with respect to the relative inter-molecular coordinate $\vec{\mathbi{r}}\/$ are denoted by the polar and azimuthal angles $\theta_{i}$ and $\phi_{i}$ (see Fig.\,\ref{fig1} for a schematic representation). The molecular rotational constant, $B_{\mathrm{rot}}$, corresponds to the rotational period $T_{\mathrm{rot}} = \hbar\pi/B_{\mathrm{rot}}$.

For the two-molecule interacting system, the field free Hamiltonian reads
\begin{equation}
\label{H}
\hat{\cal{H}} = \hat{\cal{H}}_{0} + V_{d}(\vec{\mathbi{r}})\,,
\end{equation}
where the non-interaction Hamiltonian $\hat{\cal{H}}_{0}$ is written as the following sum of mono-molecular Hamiltonians
\begin{equation}
\label{H0}
\hat{\cal{H}}_{0} = \sum_{i} \sum_{N}  ~\varepsilon_{N}~\ket{N}_{i}~\indfirst{i}{\bra{N}}\,,
\end{equation}
and the dipole interaction potential \mbox{$V_{d}(\vec{\mathbi{r}})$} takes the form
\begin{eqnarray}
\label{Vd}
V_{d}(\vec{\mathbi{r}}) & = & \frac{1}{4\pi\epsilon_0}\,\frac{\mu^2}{r^3}\,\big[ -2\cos\theta_1\cos\theta_2\nonumber\\
                      &   & \qquad +\, \sin\theta_1\sin\theta_2\cos(\phi_1-\phi_2) \,\big]\,.
\end{eqnarray}
In this equation, $\mu$ is the permanent dipole moment of the molecule.

The interaction of the two molecules with a sequence of linearly polarized laser pulses is described within the dipole approximation by the length-gauge laser interaction Hamiltonian
\begin{equation}
\label{Hl}
\hat{\cal{H}}_{\mathrm{laser}} = -\mu\,E(t) \left( \cos\theta_1 + \cos\theta_2 \right)\,,
\end{equation}
where we have assumed that the polarization of the electric field $\vec{E}(t)$ is parallel to the inter-molecular vector $\vec{\mathbi{r}}$.

Since in our scheme the projection of the rotational quantum number on the inter-molecular axis remains equal to zero, the second part of the dipole interaction potential in Eq.(\ref{Vd}) averages to zero, and one is left with
\begin{equation}
\label{Vdm=0}
V_{d}(\vec{\mathbi{r}}) = - \,\frac{1}{2\pi\epsilon_0}\,\frac{\mu^2}{r^3}\, \cos\theta_1\cos\theta_2\,.
\end{equation}
This dipole interaction only couples angular momentum states $N$ which differ by one unit, and the selection rule \mbox{$\Delta N = \pm 1$} applies for each molecule.

For the sake of simplicity let us first analyze the effect of the dipole interaction in the angular subspace spanned by the quantum numbers \mbox{$N=0$} and 1 only. This subspace is entirely characterized by the tensorial product basis set \mbox{$\ket{0}_1\otimes\ket{0}_2$}, \mbox{$\ket{0}_1\otimes\ket{1}_2$}, \mbox{$\ket{1}_1\otimes\ket{0}_2$} and \mbox{$\ket{1}_1\otimes\ket{1}_2$}, that we can reference more simply as the states \mbox{$\ket{00}$}, \mbox{$\ket{01}$}, \mbox{$\ket{10}$} and \mbox{$\ket{11}$}. Note that out of these four eigenstates of the non-interacting Hamiltonian $\hat{\cal{H}}_0$, \mbox{$\ket{01}$} and \mbox{$\ket{10}$} are degenerate. The perturbation regime therefore applies when
\begin{equation}
\label{perturb}
\left\langle 01 \left| V_{d} \right| 10 \right\rangle_{\theta_1,\theta_2} \ll 2B_{\mathrm{rot}}\,.
\end{equation}
This criterion imposes a limit on the inter-molecular separation which is discussed in Section\,\ref{sec:exp}. With such a small dipole interaction, the zero-order eigenstates of the interacting Hamiltonian $\hat{\cal{H}}$ are simply given by
\begin{subequations}
\label{psi}
\begin{eqnarray}
\psi_1 & = & \ket{00}\\
\psi_2 & = & \frac{1}{\sqrt{2}} \left( \ket{01} + \ket{10} \right)\\
\psi_3 & = & \frac{1}{\sqrt{2}} \left( \ket{01} - \ket{10} \right)\\
\psi_4 & = & \ket{11}
\end{eqnarray}
\end{subequations}
The first-order energies of $\psi_1$ and $\psi_4$ are obviously unaffected by the dipole interaction, while the degeneracy of the states \mbox{$\ket{01}$} and \mbox{$\ket{10}$} is removed at first-order, introducing the energy shifts
\begin{equation}
\label{split}
\Delta E_{\pm} = \pm \frac{1}{6\pi\epsilon_0}\,\frac{\mu^2}{r^3}\,.
\end{equation}

\begin{figure}[!t]
\includegraphics[width=8.6cm,clip]{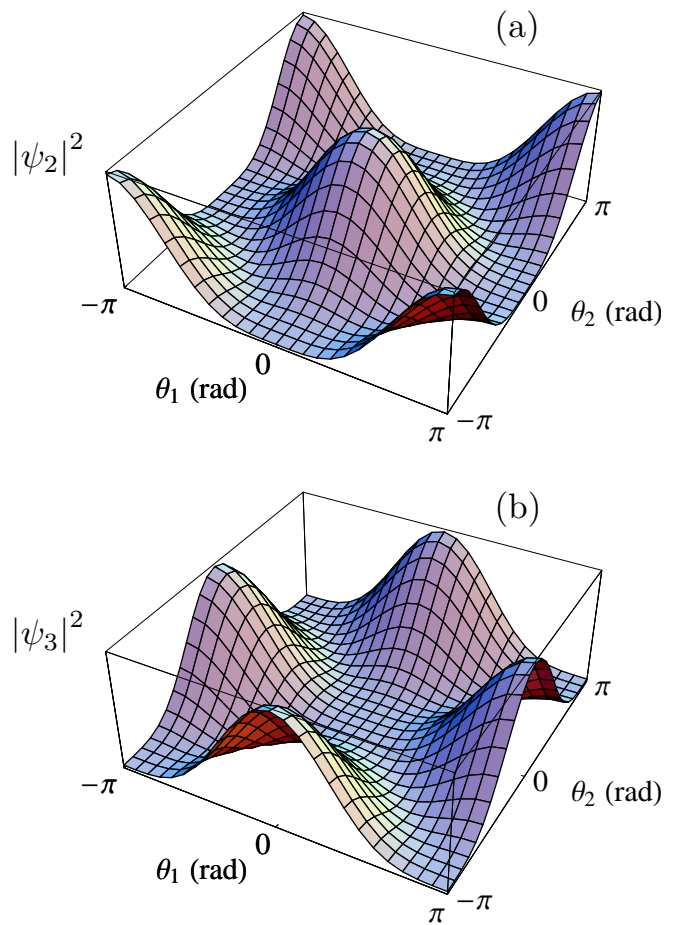}
\caption{
\label{fig:fig2}
(Color online) Probability distributions of the zero-order eigenstates of the interacting Hamiltonian $\hat{\cal{H}}$ as a function of the polar angles $\theta_1$ and $\theta_2$ defining the orientation of the two molecules with respect to the inter-molecular axis (see Fig.\,\ref{fig1})\,\cite{Note}. The upper graph~(a) corresponds to the state $\psi_2$, while the lower graph~(b) is associated with the state $\psi_3$.}
\end{figure}

The probability distribution \mbox{$P(\theta_1,\theta_2) = \left|\psi_2(\theta_1,\theta_2)\right|^2$} of state \mbox{$\psi_2$} is shown in Fig.\,\ref{fig:fig2}(a). This state, which can be seen as the following combination of molecules pointing in the same direction
\begin{equation}
\label{psi2}
\psi_2 \equiv \frac{1}{\sqrt{2}} \left( \ket{\rightarrow\rightarrow} - \ket{\leftarrow\leftarrow} \right)\,,
\end{equation}
is maximally entangled in orientation\,\cite{NOTE_ENT}. This configuration is obviously stabilized by the dipole interaction. On the other hand, the entangled state \mbox{$\psi_3$}, represented in Fig.\,\ref{fig:fig2}(b), corresponds to two molecules oriented in opposite directions
\begin{equation}
\label{psi3}
\psi_3 \equiv \frac{1}{\sqrt{2}} \left( \ket{\rightarrow\leftarrow} - \ket{\leftarrow\rightarrow} \right)\,.
\end{equation}
This state is, therefore, subjected to a repulsive interaction.

The energy shifts \mbox{$\Delta E_{\pm}$} of Eq.(\ref{split}) lead to a temporal dephasing $\Delta E_{\pm} t / \hbar$ for a free evolution during a time $t$ of the bipartite rotational states as compared to the non-interacting single-molecule rotational levels. As we will show in the following, this dephasing can be used for entanglement creation and conditional quantum logic. We now turn to the description of the mechanism we propose to perform a quantum phase gate using the two lowest rotational levels of each molecule.

\subsection{Quantum phase gate and entanglement creation}
\label{sec:QPG}

The physical implementation of quantum logic\,\cite{DIVIN} would put the predicted polynomial\,\cite{GROVER} and exponential\,\cite{SHOR} speed-up of various computational tasks of significant interest in concrete form. This achievement requires the physical implementation of a universal set of single and two-qubit quantum gates\,\cite{CHUANG}. Single-qubit operations, which consist of rotations in the qubit basis\,\cite{CHUANG}, are relatively easily implemented using stimulated Raman transitions for instance\,\cite{Revue_Masnou,PA,RbCs,Mol_OL} or using stimulated Raman adiabatic passage techniques with transform-limited laser pulses\,\cite{STIRAP}. We therefore present here a proposal for the implementation of a two-qubit quantum phase gate. This phase gate \mbox{$P(\varphi)$}, defined by the unitary transformation
\begin{equation}
\label{QPG}
\begin{array}{rcr}
\ket{00} & ~~\longrightarrow~~ & \ket{00}\\
\ket{01} & ~~\longrightarrow~~ & \ket{01}\\
\ket{10} & ~~\longrightarrow~~ & \ket{10}\\
\ket{11} & ~~\longrightarrow~~ & {\mathrm e}^{i\varphi}~\ket{11}
\end{array}
\,,
\end{equation}
entangles the two-qubits by selectively changing the state $\ket{11}$ while leaving other states unchanged. In practice, it is often simpler to implement a phase gate which changes the different qubit states according to the adiabatic transformation
\begin{equation}
\label{QPG2}
\begin{array}{rcr}
\ket{00} & ~~\longrightarrow~~ & {\mathrm e}^{i\varphi_{00}}~\ket{00}\\
\ket{01} & ~~\longrightarrow~~ & {\mathrm e}^{i\varphi_{01}}~\ket{01}\\
\ket{10} & ~~\longrightarrow~~ & {\mathrm e}^{i\varphi_{10}}~\ket{10}\\
\ket{11} & ~~\longrightarrow~~ & {\mathrm e}^{i\varphi_{11}}~\ket{11}
\end{array}
\,.
\end{equation}
This last unitary operation can then be reduced to the conditional phase gate \mbox{$P(\varphi)$} described in Eq.(\ref{QPG}), with
\begin{equation}
\label{phase}
\varphi=\varphi_{00}+\varphi_{11}-\varphi_{01}-\varphi_{10}\,,
\end{equation}
by using additional single-qubit operations\,\cite{CHUANG,LLOYD,PHASE1,PHASE2}. Its is clear that the dynamical phases acquired during the evolution of non-interacting eigenstates do not contribute to this global phase\,\cite{PHASE1}, and they will, therefore, be ignored in the following.

The case \mbox{$\varphi=\pi$} is of clear interest since this particular operation can be used to transform a separable two-qubit state into a maximally entangled state. Several different implementations of this universal gate have already been proposed or implemented with various physical systems\,\cite{VARQPG,VARQPG2}. In this study, we propose the implementation of such a conditional phase gate using both the vibrational and rotational molecular degrees of freedom.

The two rotational levels \mbox{$N=0$} and 2 of a well-defined vibrational state $v_0$ are used for the storage of information. These two states have several advantages for this purpose. First, they are easily manipulated by two-photon Raman transitions relying on an intermediate level of rotational quantum number \mbox{$N=1$}. These two-photon transitions may indeed be used to perform arbitrary one-qubit rotations. In addition, they are unaffected by the dipole interaction, which only couples, at first order, angular momentum states differing by one unit. Finally, their associated spontaneous decay rate
\begin{equation}
\label{grot}
\gamma_{\mathrm{rot}} \simeq \frac{1}{4\pi\epsilon_0}\,\frac{4\mu^2B_{\mathrm{rot}}^3}{3\hbar^4c^3}\,,
\end{equation}
corresponding to the \mbox{$(v_0,N=2) \rightarrow (v_0,N=0)$} transition, is not limiting their coherence time for the heteronuclear alkali dimers considered in this study, with \mbox{$B_{\mathrm{rot}} \simeq 0.01 - 0.1\,$cm$^{-1}$}. It is interesting to note here that, in various experiments\,\cite{Mol_OL}, cold diatomic molecules have already been prepared and trapped using optical lattices in their ground electronic and rotational levels and in a well-defined vibrational state.

\begin{figure}[!t]
\includegraphics[width=8.6cm,clip]{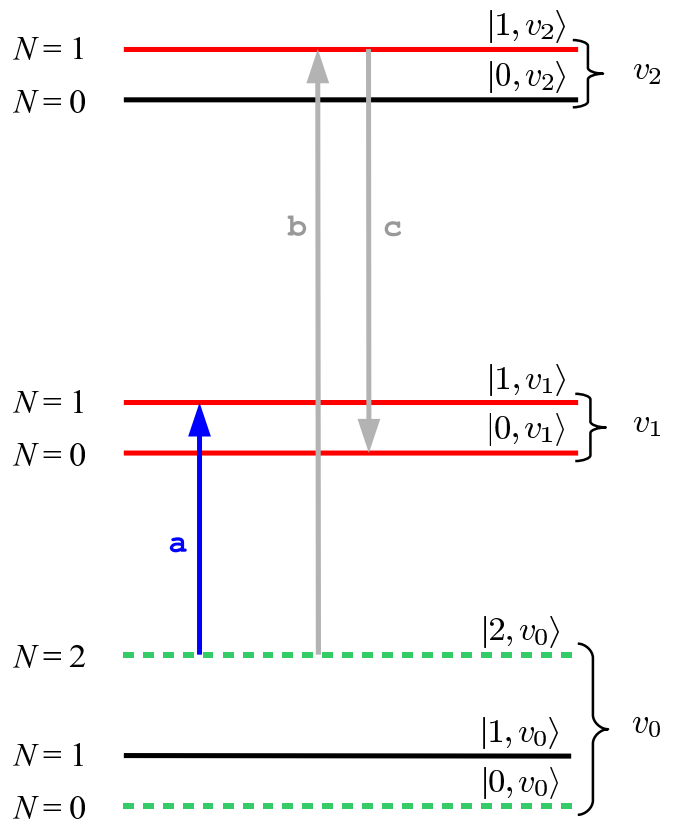}
\caption{(Color online) Laser pulses and energy levels involved in the creation of the auxiliary state \mbox{$\ket{+}$}, defined in Eq.(\ref{state+}). This transfer enables controlled dipole coupling between two neighboring molecules. The $\ket{0}$ and $\ket{1}$ qubit states are shown as two green dotted horizontal lines. The $\pi/2$~laser pulse~(a) first creates the coherent superposition \mbox{$(\ket{2,v_0}+\ket{1,v_1})/\sqrt{2}$}. The laser pulses~(b) and~(c) then transfer the remaining population of $\ket{2,v_0}$ to $\ket{0,v_1}$ in a two-photon process. Arbitrary coherent superpositions $\alpha\ket{0,v_0}+\beta\ket{2,v_0}$ can also be created from state $\ket{0,v_0}$ by a two-photon Raman process relying on a \mbox{$N=1$} intermediate level.}
\label{fig3}
\end{figure}

For the sake of simplicity, let us now denote the states of both molecules by labelling them according to the value of their rotational {\em and\/} vibrational quantum numbers as \mbox{$\ket{N,v}$}. Our qubit states are, therefore, now defined as being
\begin{subequations}
\begin{eqnarray}
\label{qubit}
\ket{0} & \equiv & \ket{0,v_0}\\
\ket{1} & \equiv & \ket{2,v_0}
\end{eqnarray}
\end{subequations}
To implement the conditional phase gate~(\ref{QPG}), the qubit state \mbox{$\ket{1}$} is selectively transferred, at time \mbox{$t=0$}, to the coherent superposition
\begin{equation}
\label{chi}
\ket{\Omega_0} = \alpha\,\ket{0,v_1} + \beta\,\ket{1,v_1}
\end{equation}
associated with a vibrational state \mbox{$v_1 \neq v_0$}, while the qubit state \mbox{$\ket{0}$} remains unchanged. The state $\ket{\Omega_0\Omega_0}$ can be easily expressed in the eigenbasis \mbox{$\left\{\psi_i\right\}$} of Eq.(\ref{psi}), where its time-evolution is simply given by analytical phase factors. We now denote the state\,(\ref{chi}) at any time \mbox{$t>0$} by \mbox{$\ket{\Omega_t}$}. A simple analysis then shows that the time-average dipole interaction
\begin{equation}
\left\langle V_{d} \right\rangle = \frac{1}{T_{\mathrm{rot}}}\int_{0}^{T_{\mathrm{rot}}} \left\langle \Omega_t\Omega_t \left| V_{d} \right| \Omega_t\Omega_t \right\rangle_{\theta_1,\theta_2} \,dt
\end{equation}
is maximized if the quantum superposition $\ket{\Omega_0}$ is chosen as the state
\begin{equation}
\label{state+}
\ket{\Omega_0} = \ket{+} = \frac{1}{\sqrt{2}}\left(\ket{0,v_1}+\ket{1,v_1}\right)\,.
\end{equation}

We will, therefore, transfer here the qubit state $\ket{1}$ to the coherent superposition \mbox{$\ket{\Omega_0}\equiv\ket{+}$}. This transfer involves two transitions, which are represented schematically with the arrows shown in Fig.\,\ref{fig3}. A first $\pi/2$~laser pulse (arrow (a) in Fig.\,\ref{fig3}) transfers half of the population from state \mbox{$\ket{1}\equiv\ket{2,v_0}$} to state \mbox{$\ket{1,v_1}$}, thus creating the coherent superposition \mbox{$(\ket{2,v_0}+\ket{1,v_1})/\sqrt{2}$}. A two-photon Raman process (arrows (b) and (c) in Fig.\,\ref{fig3}) then transfers the remaining population of state \mbox{$\ket{2,v_0}$} to state $\ket{0,v_1}$, therefore completing the protocol and generating the $\ket{+}$ state of Eq.(\ref{state+}). This transfer can be performed using stimulated Raman adiabatic passage techniques for instance\,\cite{STIRAP}. For practical reasons, it could also be preferable to operate this two-photon transition slightly detuned from the intermediate level \mbox{$\ket{1,v_2}$}. The efficiency of spontaneous Raman scattering, a mechanism possibly leading to trap losses in optical lattices, is indeed clearly decreased in this case\,\cite{Mol_OL}. In terms of pulse duration, it is necessary to work with pulses whose spectral bandwidth is much lower than the rotational energy spacing $2B_{\mathrm{rot}}$. With the polar molecules considered in this study (see Table\,\ref{table1} for the rotational constants of RbCs, KCs, KRb, NaCs, NaRb and NaK), this requirement imposes a pulse duration much larger than $\Delta t \sim 200$\,ps. Since the gate durations obtained with these polar molecules belong to the $\mu$s time scale (see Sec.\,\ref{sec:exp} hereafter), the transfer between the storage qubits and the interacting states proposed here is not limiting the total operation time of the gate. Re-establishing the initial state $\ket{2,v_0}$ is simply done by using the pulses which invert this unitary operation: a $\pi$-pulse for the transition \mbox{$\ket{1,v_1}\rightarrow\ket{2,v_0}$}, and a Raman pulse sequence similar to the one shown in Fig.\,\ref{fig3} for the complete transfer \mbox{$\ket{0,v_1}\rightarrow\ket{2,v_0}$}. Note that, due to the choice of laser frequencies, the state $\ket{2,v_0}$ is the only one affected by the laser pulses, and the state $\ket{+}$ is produced conditioned to the fact that the molecules are in state $\ket{2,v_0}$ initially. Finally, we would like to stress that, as shown in Ref.\,\cite{VARQPG2}, the fact that the interaction with the laser pulses is analyzed in terms of single-molecule states, while the overall two-qubit phase gate operation is based on the two-molecule interacting Hamiltonian, is not limiting the generality of our results.

When this sequence of laser pulses is applied to both molecules simultaneously, the complete molecular system ends up in state $\ket{++}$ if it was initially in state $\ket{11}$.

\begin{figure}[!t]
\includegraphics[width=8.6cm,clip]{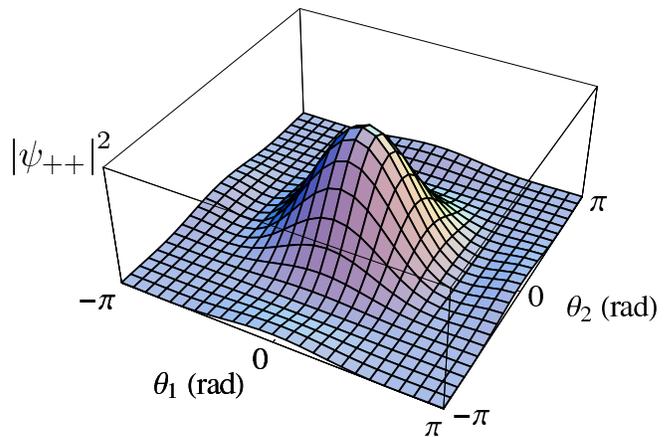}
\caption{
\label{fig:fig4}
(Color online) Probability distribution associated with the interacting state \mbox{$\psi_{++}\equiv\ket{++}$} corresponding to both molecules in the $\ket{+}$ state~(\ref{state+}) as a function of the polar angles $\theta_1$ and $\theta_2$ which define the orientation of the two molecules with respect to the inter-molecular axis (see Fig.\,\ref{fig1})\,\cite{Note}.}
\end{figure}

The probability distribution of state $\ket{++}$ is shown in Fig.\,\ref{fig:fig4} as a function of $\theta_1$ and $\theta_2$. This separable state clearly corresponds to two molecules oriented in the same direction, with
\begin{equation}
\ket{++}\equiv\ket{\rightarrow\rightarrow}\,.
\end{equation}
Since the two coherent superpositions~(\ref{state+}) which are associated with each molecule evolve in phase, this parallel orientation is maintained at any time. As a consequence, this state is stabilized by the dipole interaction.

This stabilization effect can also be easily deduced from a simple rewriting of state $\ket{++}$ as a function of the eigenstates~(\ref{psi}) of the interacting Hamiltonian $\hat{\cal{H}}$
\begin{equation}
\label{state++}
\ket{++} = \frac{1}{2}\left(\psi_1+\psi_4\right)+\frac{1}{\sqrt{2}}\,\psi_2\,.
\end{equation}
This rewriting reveals the contribution of the stabilized eigenstate~$\psi_2$, and a lack of contribution from the destabilized state~$\psi_3$.

The quantum phase gate can, therefore, be implemented in three steps, following the sequence
\begin{equation}
\label{QPG3}
\begin{array}{rccclcr}
\ket{00} & \longrightarrow & \ket{00} & \longrightarrow &  \quad\ket{00} & \longrightarrow &  \ket{00}\\
\ket{01} & \longrightarrow & \ket{0+} & \longrightarrow &  \quad\ket{0+} & \longrightarrow &  \ket{01}\\
\ket{10} & \longrightarrow & \ket{+0} & \longrightarrow &  \quad\ket{+0} & \longrightarrow &  \ket{10}\\
\ket{11} & \longrightarrow & \ket{++} & \longrightarrow & -\ket{++} & \longrightarrow & -\ket{11}
\end{array}
\,,
\end{equation}
where the dynamical phases associated with the non-interacting evolution have been ignored. The sign change of state $\ket{++}$ is due to the free evolution of the dipole interacting two-molecule system. Indeed, because of the dipole interaction, this state gains a time dependent phase easily expressed as $\varphi(t) = (\delta/\hbar)\,t$, where
\begin{equation}
\label{E_shift}
\delta  =  \left\langle ++ \left| V_d \right| ++ \right\rangle
=\frac{1}{4\pi\epsilon_0}\,\frac{\mu^2}{3 r^3}
\end{equation}
This step is, therefore, able to build up a maximal molecular entanglement in the duration
\begin{equation}
\label{tau}
\tau=4\pi\epsilon_0\left(\frac{3 \hbar\pi r^3}{\mu^2}\right)
\end{equation}
from an initially separable two-qubit wavefunction.

Since, from all accessible states in Eq.(\ref{QPG3}), the state $\ket{++}$ is the only one in which the two molecules are coupled by dipole interaction at first order, by transforming the $\ket{+}$ states back to the ``storage'' qubit states $\ket{1}$, we stop the conditional interaction and transfer the quantum phase gate to the original subspace spanned by the levels $\ket{00}$, $\ket{01}$, $\ket{10}$ and $\ket{11}$.

It is important to notice that the quantum numbers $M_i$, projections of the rotational quantum numbers of the two molecules on the inter-molecular axis, remain unchanged in the protocol above. This happens thanks to the polarization chosen for the electric field \mbox{$\vec{E}(t)$} (see Fig.\,\ref{fig1}), and because the dipole interaction~(\ref{Vdm=0}) can be expressed as the $Y_{2,0}$ component of a second order spherical tensor. This interaction therefore conserves the total projection \mbox{${\cal{M}}=M_1+M_2$}. In our scheme, the initial value of ${\cal{M}}$ is zero, and the dipole coupling affects the linear combination \mbox{$\ket{01}+\ket{10}$} of state $\ket{++}$ only. In this linear combination, one of the molecule remains in the ground rotational level \mbox{$N=0$}, and both projections $M_1$ and $M_2$ are, therefore, fixed at zero during the whole gate duration.

\begin{table}[!t]
\caption{
\label{table1}
Optimal range of inter-molecular separations for the implementation of conditional quantum logic with polar molecules. The minimum and maximum distances $r_{\mathrm{min}}$ and $r_{\mathrm{max}}$ verify the inequalities (\ref{perturb2}) and (\ref{decoherence}), with a ratio between the left and right hand sides of these equations equal to 10$^3$. The molecular parameters $B_{\mathrm{rot}}$, $\mu$ and $\omega_{\mathrm{vib}}$ are taken from\,\cite{Aymar} and\,\cite{Herzberg}.
}
\begin{ruledtabular}
\begin{tabular}{cccccc}
 & $B_{\mathrm{rot}}$ & $\mu$ & $\omega_{\mathrm{vib}}$ & $r_{\mathrm{min}}$ & $r_{\mathrm{max}}$\\
 & (cm$^{-1}$)        & (D)   & (cm$^{-1}$)             & (nm)               & (nm)\\
\hline
RbCs & 1.65 10$^{-2}$ & 1.21 &  49.4 & 52.8 & 1385 \\
KCs  & 3.08 10$^{-2}$ & 1.84 &  66.2 & 56.8 & 1033 \\
KRb  & 3.80 10$^{-2}$ & 0.59 &  75.5 & 24.8 &  906 \\
NaCs & 5.88 10$^{-2}$ & 4.58 &  98.0 & 84.3 &  698 \\
NaRb & 7.02 10$^{-2}$ & 3.30 & 107.0 & 63.8 &  639 \\
NaK  & 9.81 10$^{-2}$ & 2.76 & 124.1 & 50.7 &  551 \\
\end{tabular}
\end{ruledtabular}
\end{table}

Finally, we also would like to stress that, in the protocol described above, the auxiliary vibrational levels could as well be replaced by electronic states without substantially modifying the basic procedure of entanglement creation. In this case, it would however be required to implement the two-photon transition shown in Fig.\,\ref{fig3} slightly detuned from the intermediate level in order to avoid spontaneous emission by electric dipole transitions\,\cite{Mol_OL}.

\section{Discussion on possible experimental implementations}
\label{sec:exp}

A practical implementation of the quantum phase gate described in Section\,\ref{sec:QPG} must satisfy the following conditions:
\begin{itemize}
\item[(i)]
the molecules should be prepared initially in their rotational ground state,
\item [(ii)]
the molecules should be close enough for a fast operation time,  but far enough to avoid strong non-linear interactions,
\item [(iii)]
the molecules should be individually addressable,
\item [(iv)]
the inter-molecular distance should remain constant during the whole gate duration,
\item [(v)]
the gate operation time should be much shorter than the decoherence time.
\end{itemize}

Considering the five requirements above, trapped molecular systems seem to be interesting candidates for the implementation of the quantum phase gate protocol. We proceed now to a more quantitative investigation of the experimental parameters involved.

Let us take as an example cold molecules trapped in an optical lattice. In such systems, the molecules can be formed from an atomic Bose-Einstein condensate (BEC) by Feshbach resonance or by photoassociation. Since the molecules are then formed in $s$-wave collisions, the first criterion is necessarily fulfilled. In addition, the molecules could be addressed individually (criterion (iii)) as proposed by DeMille\,\cite{DEMILLE} by using an electric field gradient which shifts the transition frequencies of the different molecules as a function of their position. Finally, when the dipole interaction is weak, the molecules are not moving from their lattice site during the gate operation, and the criterion (iv) is also verified.

For the dipole interaction to be treated as a perturbation, one should require that the inequality~(\ref{perturb}) is verified. The two molecules should therefore be well separated, with
\begin{equation}
\label{perturb2}
r^3\; \simeq \;\left(\frac{\lambda}{2}\right)^{\!\!3}\; \gg \;\frac{1}{4\pi\epsilon_0}\left(\frac{\mu^2}{3 B_{\mathrm{rot}}}\right)\,.
\end{equation}
In this equation, $\lambda$ denotes the wavelength of the lattice laser light.

A link can be made between this requirement and the condition~(v). In an optical lattice, among various sources of decoherence, one can cite spontaneous emission, decoherence due to the coupling to the black-body radiation of the room-temperature environment, and collisions with residual atoms or molecules.  The black-body contribution leads to lifetimes which are much larger \mbox{($\geqslant 100\,$s)} than the expected gate duration\,\cite{KOTO,KOTO2}. For such cold and relatively isolated molecules, spontaneous emission from excited vibrational states should therefore present the highest contribution to decoherence. The spontaneous vibrational decay rate is then given by
\begin{equation}
\gamma_{\mathrm{vib}} \simeq \frac{1}{4\pi\epsilon_0}\left(\frac{4\mu^2\omega_{\mathrm{vib}}^3}{3\hbar c^3}\right)\,,
\end{equation}
where $\omega_{\mathrm{vib}}$  is the vibrational frequency and $c$ is the speed of light. The requirement~(v), reformulated as 
\begin{equation}
\gamma_{\mathrm{vib}}\,\tau \ll 1\,,
\end{equation}
therefore yields another criterion for the inter-molecular separation
\begin{equation}
\label{decoherence}
r^3\; \simeq \;\left(\frac{\lambda}{2}\right)^{\!\!3}\; \ll \;\frac{c^3}{4\pi\omega_{\mathrm{vib}}^3}\,.
\end{equation}

Combining Eqs.~(\ref{perturb2}) and~(\ref{decoherence}) yields an optimal range \mbox{[\,$r_{\mathrm{min}}$\,,\,$r_{\mathrm{max}}$\,]} of inter-molecular separations for the implementation of the present conditional quantum logic scheme with polar molecules. This range of distances is given in Table\,\ref{table1} for a set of alkali dimers which have been presented as potential candidates for quantum information\,\cite{DEMILLE,LEE,KOTO}.

Clearly, all alkali dimers are well suited if one considers lattices in the optical or near infrared domain, around \mbox{$\lambda \simeq 600-1100\,$nm}. In addition, a recent study\,\cite{KOTO} has shown the existence of two frequency windows for KRb and RbCs in which, in spite of the complex molecular internal structure, resonant excitation by the lattice light is very unlikely. The trapping potential of the lattice is then almost unaffected by this additional complexity.

\begin{figure}[!t]
\includegraphics[width=8.6cm,clip]{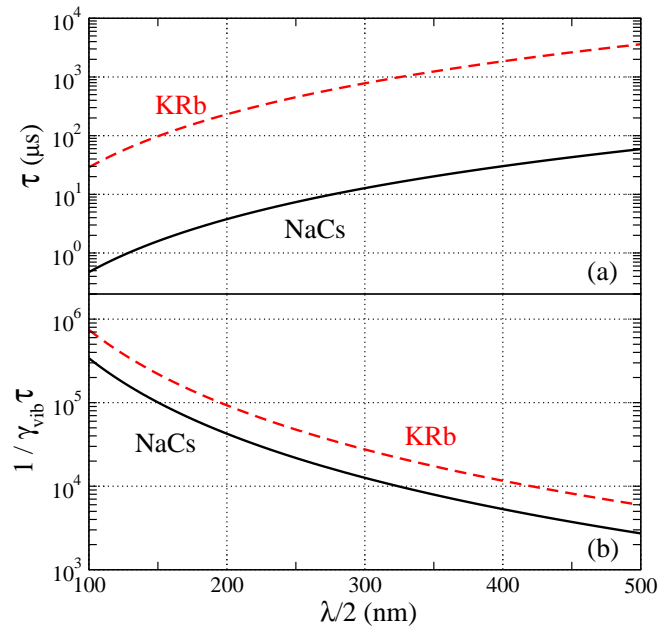}
\caption{
\label{fig:fig5}
(Color online) Gate duration $\tau$ [upper panel (a)] and gate robustness $1/\gamma_{\mathrm{vib}}\tau$ [lower panel (b)] in logarithmic scales as a function of the average inter-molecular separation $r\simeq\lambda/2$ in an optical lattice of wavelength $\lambda$. The values corresponding to NaCs ($\mu=4.58\,$D) are shown as black solid lines and the one associated with KRb ($\mu=0.59\,$D) are represented as red dashed lines.}
\end{figure}

The efficiency of the present entanglement procedure is finally analyzed in Fig.\,\ref{fig:fig5} for the two dimers KRb and NaCs, which have the smallest and largest dipole moments \mbox{$\mu=0.59\,$D} and \mbox{$\mu=4.58\,$D} of the alkali molecules of Table\,\ref{table1}. The upper and lower panels (a) and (b) of this figure show the expected gate duration $\tau$ (Eq.(\ref{tau})) and the gate robustness \mbox{$1/\gamma_{\mathrm{vib}}\tau$} as a function of the lattice site separation \mbox{$\lambda/2$} for these two molecules.

In the domain \mbox{$\lambda/2 \simeq 300-500\,$nm} (right part of Fig.\,\ref{fig:fig5}), gate durations around \mbox{$\tau \simeq 12-60\,\mu$s} and \mbox{$\tau \simeq 0.8-3.6\,$ms} are obtained for NaCs and KRb respectively. As shown in Fig.\,\ref{fig:fig5}(b), in the same range of wavelengths, about $3 \times 10^3$ to $3 \times 10^4$ gates can be performed in the expected coherence time of the system. More specifically, the frequency window \mbox{$\lambda \simeq 680 \pm 35\,$nm} given in Ref.\,\cite{KOTO} for KRb yields the gate duration \mbox{$\tau \simeq 1.2\,$ms}, with \mbox{$1/\gamma_{\mathrm{vib}}\tau \simeq 1.8\,10^4$} gates achievable in the expected coherence time. In addition, one should note that the gate duration $\tau$ is much larger than the molecular rotational period. The quantum phase gate therefore builds up in a very large number of molecular rotations.

The number of rotations necessary for the creation of entanglement can be decreased by using molecules closer to each other. This can be achieved in systems where the molecular density reaches higher values, as in solid matrices for instance. In such systems, the quantum phase gate protocol described above leads to {\it uncontrolled} entanglement creation between different molecules. Since in this system the molecules are randomly located in the matrix sites, one can usually not define rigorously a unique inter-molecular quantization axis. As a consequence, different values of ${\cal{M}}$ will appear during the implementation of the quantum phase gate protocol. For relatively low densities, an undetermined entangled state is, therefore, produced between pairs of molecules.

Let us now consider the specific case of DCl \mbox{($\mu \simeq 1.02\,$D)} trapped in an fcc Ar crystal\,\cite{CLAUDINE}, with an Ar-Ar distance equal to that in bulk Ar, {\em i.e.\/} \mbox{7.03\,a.u.}. The DCl molecules are located at the center of the cubic Ar structure, and the closest molecules are separated from each other by \mbox{7.03\,a.u.} only. In these conditions, a quantum phase gate may be performed in just \mbox{$\tau \simeq 500\,$fs}.

The vibrational motion of these molecules is of course coupled to the vibrational modes of the crystal, and this is the main source of vibrational decoherence in this system. This decoherence time strongly depends on temperature, and for \mbox{$T=6\,$K}, it is of the order of 100\,ps\,\cite{CLAUDINE}. About 200 gates could therefore be performed within the system coherence time. By further cooling down this system, one can expect to dramatically increase these figures. An advantage of this type of system is its simplicity, and the fact that it is a tractable macroscopic system. It also allows, in principle, for a direct test of entanglement, as discussed in the next Section.

\section{Entanglement detection and non-locality tests}
\label{sec:detect}

A possible way to completely characterize the entangled state generated by the procedure described in Section\,\ref{sec:creation} is to perform a tomographic measurement of the rotational state. This can be done by detecting, by measurements of molecular alignment for instance \cite{Dooley}, the complete density matrix of the system\,\cite{MOLMER}. This procedure, usually employed in the quantum information context with trapped atoms and ions, presents the advantage of giving complete information about the system state.

However, if one is interested only in determining if both molecules are entangled or not, other measurement schemes can be simpler and more direct. They are based on {\it entanglement witnesses}\,\cite{WITNESS}. These measurements present the disadvantage of not providing necessary and sufficient criteria for entanglement detection since only a subspace of the Hilbert space spanned by all entangled states is detected.

A possible way of testing entanglement using entanglement witnesses is to perform  Bell type experiments\,\cite{BELL_CHSH} with molecules in an optical lattice. By applying the unitary transformations which allow for the implementation of a conditional phase gate between the storage rotational levels $\ket{0}$ and $\ket{1}$ as discussed in Section\,\ref{sec:creation}, and then by measuring the population of each level, one can infer the quantity
\begin{equation}
\label{bell}
{\cal B}=\big| \langle \sigma_a \sigma_b \rangle + \langle \sigma_{a'} \sigma_b \rangle + \langle \sigma_{a} \sigma_{b'} \rangle - \langle \sigma_{a'} \sigma_{b'} \rangle \big|\,,
\end{equation}
where the $\sigma_{\!\alpha}$ are the Pauli matrices in the $\alpha$ directions, written in the \mbox{$\left(\ket{0},\ket{1}\right)$} basis set. In this type of measurement, the molecules do not need to be distinguished. As shown in Ref.\,\cite{BELL_CHSH}, all separable states satisfy the inequality
\begin{equation}
{\cal B} \leqslant 2\,,
\end{equation}
while some entangled states violate it. Other types of approaches are also possible with temporal Bell inequalities, for instance\,\cite{Or_ent}.

In the solid matrix system, the rotational states of the molecules are not as accessible as in the preceding case. In such systems, the measurements are usually made by radiation detection or by photon echo techniques\,\cite{CLAUDINE}. In Ref.\,\cite{EU}, it was shown that photon echoes can be used as entanglement witnesses, detecting a subspace of entangled states. In the experimental system in question, the echoes are observed in a transition between two vibrational levels. In order to obtain information about entanglement between rotational states, one should first coherently transfer the population of state $\ket{1}\equiv\ket{2,v_0}$ to the first excited vibrational state with no rotational excitation, $\ket{0,v_1}$, using the two-photon process shown in Fig.\,\ref{fig3}. The result of this transformation is somehow to swap the excitations associated with the rotational and vibrational degrees of freedom. By doing so, entanglement between rotational levels is, therefore, converted into entanglement between vibrational levels. The techniques described in\,\cite{EU} can then be applied for entanglement detection.

\section{Conclusion}

We have presented an alternative way for creating controlled and uncontrolled entanglement between rotational levels of two polar diatomic molecules trapped in optical lattices or in solid matrices.

Our scheme is based on a weak dipole coupling which is conditionally created between molecules, leading to a conditional phase shift. It uses the three lowest rotational levels of two vibrational states. For storage of the information between the gate operations, the qubit state is transferred efficiently via a Raman transition to two uninteracting states of long coherence times.

We have discussed two possible experimental scenarios which are suitable for implementing the proposed scheme, as well as possible detection techniques adapted to both experimental contexts. These results throw some light on how to perform quantum information operations in cold and trapped molecular systems.

\begin{acknowledgments}
The authors are indebted to Claudine Cr\'epin and Michel Broquier for inspiring discussions. During the realization of this work, P\'erola Milman was financially supported by the ACI {\it ``Molecular Simulations"}. Eric Charron acknowledges financial support from CEA (contract number LRC-DSM 05-33) and HPC facilities of IDRIS-CNRS (contract number 05-1848). Laboratoire de Photophysique Mol\'eculaire is associated with Universit\'e Paris-Sud.
\end{acknowledgments}



\begin{thebibliography}{99}

\bibitem{HAROCHE}
A. Rauschenbeutel, G. Nogues, S. Osnaghi, P. Bertet, M. Brune, J. M. Raimond, and S. Haroche, Science {\bf 288}, 2024 (2000) ;
J. M. Raimond, M. Brune, and S. Haroche, \rmp {\bf 73}, 565 (2001).

\bibitem{BLATT_WINE}
H. H\"affner, W. H\"ansel, C. F. Roos, J. Benhelm, D. Chek-al-kar, M. Chwalla, T. K\"orber, U. D. Rapol, M. Riebe, P. O. Schmidt, C. Becher, O. G\"uhne, W. D\"ur and R. Blatt, Nature {\bf 438}, 643 (2005) ;
D. Leibfried, E. Knill, S. Seidelin, J. Britton, R. B. Blakestad, J. Chiaverini, D. B. Hume, W. M. Itano, J. D. Jost, C. Langer, R. Ozeri, R. Reichle and D. J. Wineland, Nature {\bf 438}, 639 (2005).

\bibitem{NMR}
L. M. K. Vandersypen and I. L. Chuang, \rmp {\bf 76}, 1037 (2004).

\bibitem{NA}
O. Mandel, M. Greiner, A. Widera, T. Rom, T. W. H\"ansch, and I. Bloch, Nature {\bf 425}, 937 (2003) ;
J. V. Porto, S. Rolston, B. Laburthe Tolra, C. J. Williams, and W. D. Phillips, Phil. Trans. R. Soc. Lond. A {\bf 361}, 1417 (2003) ;
D. Jaksch, Cont. Phys. {\bf 45}, 367 (2004).

\bibitem{Revue_Masnou}
Special Issue, {\em Ultracold Polar Molecules: Formation and Collisions\/}, Edited by J. Doyle, B. Friedrich, R.V. Krems, and F. Masnou-Seeuws, Eur. Phys. J. D {\bf 31}, 149 (2004).

\bibitem{PA}
R. Wymar, R. S. Freeland, D. J. Han, C. Ryu, and D. J. Heinzen, Science {\bf 287}, 1016 (2000) ;
C. McKenzie, J. HeckerDenschlag, H. H\"affner, A. Browaeys, L. E. E. de Araujo, F. K. Fatemi, K. M. Jones, J. E. Simsarian, D. Cho, A. Simoni, E. Tiesinga, P. S. Julienne, K. Helmerson, P. D. Lett, S. L. Rolston, and W. D. Phillips, \prl {\bf 88}, 120403 (2002).

\bibitem{FR}
E. A. Donley, N. R. Claussen, S. T. Thompson, and C. E. Wieman, Nature {\bf 417}, 529 (2002) ;
C. A. Regal, C. Ticknor, J. L. Bohn, and D. S. Jin, Nature {\bf 424}, 47 (2003) ;
J. Herbig, T. Kraemer, M. Mark, T. Weber, C. Chin, H. C. N\"agerl, and R. Grimm, Science {\bf 301}, 1510 (2003).

\bibitem{RbCs}
A. J. Kerman, J. M. Sage, S. Sainis, T. Bergeman, and D. DeMille, \prl {\bf 92}, 033004 (2004) ;
J. M. Sage, S. Sainis, T. Bergeman, and D. DeMille, \prl {\bf 94}, 203001 (2005).

\bibitem{LiNa}
C. A. Stan, M. W. Zwierlein, C. H. Schunck, S. M. F. Raupach, and W. Ketterle, \prl {\bf 93}, 143001 (2004).

\bibitem{KRb}
M. W. Mancini, G. D. Telles, A. R. L. Caires, V. S. Bagnato, and L. G. Marcassa, \prl {\bf 92}, 133203 (2004) ;
S. Inouye, J. Goldwin, M. L. Olsen, C. Ticknor, J. L. Bohn, and D. S. Jin, \prl {\bf 93}, 183201 (2004) ;
D. Wang, J. Qi, M. F. Stone, O. Nikolayeva, H. Wang, B. Hattaway, S. D. Gensemer, P. L. Gould, E. E. Eyler, and W. C. Stwalley, \prl {\bf 93}, 243005 (2004).

\bibitem{NaCs}
C. Haimberger, J. Kleinert, M. Bhattacharya, and N. P. Bigelow, \pra {\bf 70}, 021402(R) (2004).

\bibitem{Mol_OL}
T. Rom, T. Best, O. Mandel, A. Widera, M. Greiner, T. W. H\"ansch, and I. Bloch, \prl {\bf 93}, 073002 (2004) ;
T. St\"oferle, H. Moritz, K. G\"unter, M. K\"ohl, and T. Esslinger, \prl {\bf 96}, 030401 (2006) ;
G. Thalhammer, K. Winkler, F. Lang, S. Schmid, R. Grimm, and J. H. Denschlag, \prl {\bf 96}, 050402 (2006).

\bibitem{H_Mol_OL}
M. G. Moore and H. R. Sadeghpour, \pra {\bf 67}, 041603(R) (2003) ;
T. Miyakawa and P. Meystre, \pra {\bf 73}, 021601(R) (2006).

\bibitem{ERIC}
E. Charron and M. Raoult, \pra {\bf 74}, 033407 (2006).

\bibitem{MOLQI}
Z. Amitay, R. Kosloff, and S. R. Leone, Chem. Phys. Lett. {\bf 359}, 8 (2002) ;
J. Vala, Z. Amitay, B. Zhang, S. R. Leone, and R. Kosloff, \pra {\bf 66}, 062316 (2002).

\bibitem{MOLQI2}
J. P. Palao, and R. Kosloff, \prl {\bf 89}, 188301 (2002).

\bibitem{MOLQI3}
C. M. Tesch, and R. de Vivie-Riedle, \jcp {\bf 121}, 12158 (2004).

\bibitem{ZOLLER}
A. Micheli, G. K. Brennen, and P. Zoller, Nature Phys. {\bf 2}, 341 (2006).

\bibitem{DEMILLE}
D. DeMille, \prl {\bf 88}, 067901 (2002).

\bibitem{LEE}
C. Lee and E. A. Ostrovskaya, \pra {\bf 72}, 062321 (2005).

\bibitem{KOTO}
S. Kotochigova and E. Tiesinga, \pra {\bf 73}, 041405(R) (2006).

\bibitem{Yelin}
S. F. Yelin, K. Kirby, and R. C\^ot\'e, \pra {\bf 74}, 050301(R) (2006).

\bibitem{SCIENCE}
C. Hettich, C. Schmitt, J. Zitzmann, S. K\"uhn, I. Gerhardt, and V. Sandoghdar, Science {\bf 298}, 385 (2002).

\bibitem{KOTO2}
E. Tiesinga, S. Kotochigova, and P. S. Julienne, \pra {\bf 65}, 042722 (2002) ;
S. Kotochigova and E. Tiesinga, \jcp {\bf 123}, 1 (2005).

\bibitem{Note}
Even though the polar angle $\theta$ is defined on the interval \mbox{[0,$\pi$]} only, the probability distributions shown in this graph are periodically extended to the interval \mbox{[-$\pi$,$\pi$]} in order to improve the legibility of the figure.

\bibitem{NOTE_ENT}
For bipartite pure states, the von Neumann entropy associated with one of the subsystems provides an unambiguous measure of the degree of entanglement (see chapter\,11 of Ref.\,\cite{CHUANG} for details). In a $d-$dimensional Hilbert space and for bipartite pure states, a maximal entanglement is obtained when this entropy reaches its maximum value $\log_2(d)$.

\bibitem{DIVIN}
P. Divincenzo, Fortschr. Phys. {\bf 48}, 771 (2000).

\bibitem{GROVER}
L. K. Grover, {\em Proceedings of the 28th Annual ACM Symposium on the Theory of Computing\/}, p. 212, (1996), and quant-ph/9605043 ;
L. K. Grover, \prl {\bf 79}, 325 (1997).

\bibitem{SHOR}
P. W. Shor, {\em Proc. 5th Annual Symp. on Found. of Comput. Science\/}, IEEE Comput. Soc. Press (1994) ;
P. W. Shor, SIAM J. Sci. Statist. Comput. {\bf 26} 1484 (1997), and quant-ph/9508027.

\bibitem{CHUANG}
M. A. Nielsen and I. L. Chuang, {\em Quantum Computation and Quantum Information\/}, Cambridge University Press, Cambridge (2000).

\bibitem{STIRAP}
U. Gaubatz, P. Rudecki, S. Schiemann, and K. Bergmann, \jcp {\bf 92}, 5363 (1990) ;
K. Bergmann, H. Theuer, and B. W. Shore, \rmp {\bf 70}, 1003 (1998).

\bibitem{LLOYD}
S. Lloyd, \prl {\bf 75}, 346 (1995).

\bibitem{PHASE1}
T. Calarco, E. A. Hinds, D. Jaksch, J. Schmiedmayer, J. I. Cirac, and P. Zoller, \pra {\bf 61}, 022304 (2000).

\bibitem{PHASE2}
E. Charron, E. Tiesinga, F. Mies, and C. Williams, \prl {\bf 88}, 077901 (2002) ;
D. Vager, B. Segev, and Y. B. Band, \pra {\bf 72}, 022325 (2005).

\bibitem{VARQPG}
A. Rauschenbeutel, G. Nogues, S. Osnaghi, P. Bertet, M. Brune, J.M. Raimond, and S. Haroche, \prl {\bf 83}, 5166 (1999) ;
E. Charron, E. Tiesinga, F. Mies, and C. Williams, \prl {\bf 88} 077901 (2002) ;
N. Kiesel, C. Schmid, U. Weber, R. Ursin, and H. Weinfurter, \prl {\bf 95}, 210505 (2005).

\bibitem{VARQPG2}
M. A. Cirone, A. Negretti, T. Calarco, P. Krüger, and J. Schmiedmayer, Eur. Phys. J. D 35, 165 (2005);
E. Charron, M. A. Cirone, A. Negretti, J. Schmiedmayer, and T. Calarco, \pra {\bf 74}, 012308 (2006).


\bibitem{Aymar}
M. Aymar and O. Dulieu, \jcp {\bf 122}, 204302 (2005).

\bibitem{Herzberg}
G. Herzberg, {\em Molecular Spectra and Molecular Structure, I.Spectra of Diatomic Molecules\/}, 2nd ed., D. Van Nostrand Co., NY (1950).

\bibitem{CLAUDINE}
M. Broquier, C. Cr\'epin, A. Cuissot, H. Dubost, and J. P. Galaup, Eur. Phys. J. D {\bf 36}, 41 (2005)  ;
M. Broquier, C. Cr\'epin, A. Cuisset, H. Dubost, J. P. Galaup, and P. Roubin, J. Chem. Phys. {\bf 118}, 9582 (2003) ;
C. Cr\'epin, M. Broquier, H. Dubost, J. P. Galaup, J. L. Le Gou\"et, and J. M. Ortega, \prl {\bf 85}, 964 (2000).

\bibitem{Dooley}
 P. W. Dooley, I. V. Litvinyuk, K. F. Lee, D. M. Rayner, M. Spanner, D. M. Villeneuve, and P. B. Corkum, \pra {\bf 68}, 023406 (2003).

\bibitem{MOLMER}
A. S. Mouritzen and K. M\o{}lmer, \jcp {\bf 124}, 244311 (2006).

\bibitem{WITNESS}
B. Terhal, Phys. Lett. A {\bf 271}, 319 (2000) ;
D. Bruss, J. Math. Phys. {\bf 43}, 4237 (2002).

\bibitem{BELL_CHSH}
J. S. Bell, Physics (Long Island Ciy, New York) {\bf 1}, 195 (1964) ;
J. F. Clauser, M. A. Horne, A. Shimony, and R. A. Holt, \prl {\bf 23}, 880 (1969).

\bibitem{Or_ent}
P. Milman, A. Keller, E. Charron, and Osman Atabek, quant-ph/0612044 (2006).

\bibitem{EU}
P. Milman, \pra {\bf 74}, 042317 (2006).

\end{thebibliography}
\end{document}